\documentclass[twocolumn,letter]{emulateapj}
\usepackage{graphicx}
\usepackage{latexsym,amssymb}
\usepackage{amsmath}
\usepackage{natbib}

\def\kms{$\mbox{km s}^{-1}$}
\def\kmsmpc{$\mbox{km s}^{-1}\mbox{ Mpc}^{-1}$}

\slugcomment{\today}

\shorttitle{On a method to resolve the nuclear activity in galaxies} 

\shortauthors{Lindblad et al.}

\begin{document}

\title{On a method to resolve the nuclear activity in galaxies as applied to the Seyfert~2 galaxy NGC~1358$^{\star}$}

\author{Per Olof Lindblad$^{1}$,
Kambiz Fathi$^{1,2}$,
Maja Hjelm$^{1}$,
Charles H. Nelson$^{3}$}
\altaffiltext{1}{Stockholm Observatory, Department of Astronomy, Stockholm University, AlbaNova, 106 91 Stockholm, Sweden}
\altaffiltext{2}{Oskar Klein center for Cosmoparticle Physics, Stockholm University, 106 91 Stockholm, Sweden}
\altaffiltext{3}{Physics and Astronomy Department, Drake University, 2507 University Avenue, Des Moines, IA 50311, USA}
\altaffiltext{$\star$}{Based on observations at the European Southern Observatory.}

\begin{abstract}
Nuclear regions of galaxies generally host a mixture of
components with different excitation, composition, and kinematics.
Derivation of emission line ratios and kinematics could then be
misleading, if due correction is not made for the limited spatial
and spectral resolutions of the observations.
The aim of this paper is to demonstrate, with application to a
long slit spectrum of the Seyfert~2 galaxy NGC~1358, how line
intensities and velocities, together with modelling and knowledge
of the point spread function, may be used to resolve the differing
structures.
In the situation outlined above, 
the observed kinematics differs for different spectral lines.
From the observed intensity and velocity distributions of a
number of spectral lines, and with some reasonable assumptions 
about structure of different subcomponents 
to diminish the number of free parameters, the true line ratios
and velocity structures may be deduced. 
A preliminary solution for the nuclear structure of NGC~1358
is obtained, involving a nuclear point source and an emerging
outflow of high excitation, ending with shock and postshock cloud 
as revealed by the 
velocities, as well as a nuclear emission
line disk rotating in the potential of a stellar bulge and 
expressing a radial excitation gradient.
The method results in a likely scenario for the nuclear structure
of the Seyfert~2 galaxy NGC~1358. For definitive results an
extrapolation of the method to two dimensions combined with the 
use of integral field spectroscopy will be necessary.
\end{abstract}

\keywords{ methods: data analysis -- 
	 galaxies: structure -- 
	 galaxies: kinematics and dynamics --
	 galaxies: active}

\maketitle

\section{Introduction}
The nuclear and circumnuclear activity of galaxies generally involves the
interplay between a number of different components and phenomena, e.g. a
central active source surrounded by an absorbing torus, a rotating central
bulge, outflowing jets, flowing streams due to the action of a bar, or even
merging. To separate these different components and derive their respective
line ratios and kinematic behaviour is generally difficult due to the limited
spatial and spectroscopic resolution available. On the other hand, such a
separation is crucial to the analysis of the structures and physical processes
involved in the nuclear region and their roles in galaxy evolution.

Then, as to be demonstrated here, such a separation could benefit by considering the
differences of the velocities observed for different spectral lines and be eased
by models of the activity, smoothed with the point spread function (PSF) and fitted
to the observations. 

\begin{figure}
\resizebox{\hsize}{!}{\includegraphics{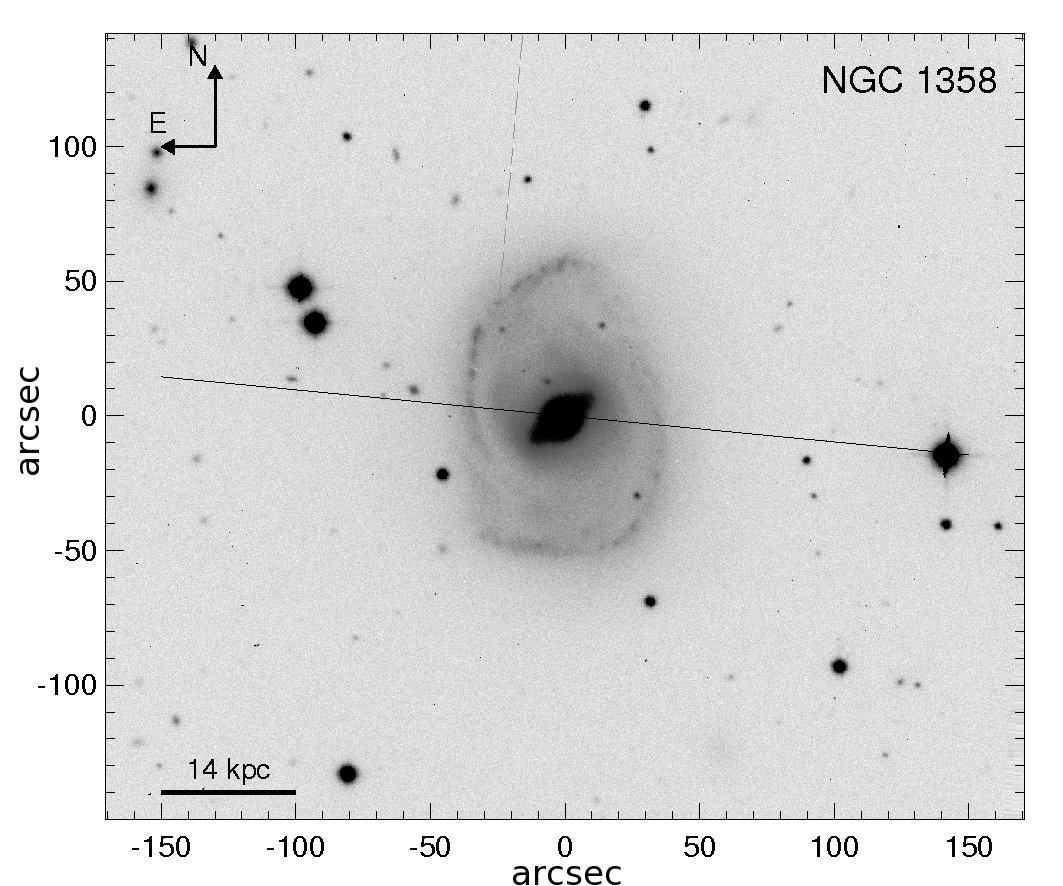}}
\caption{NGC~1358 optical image obtained with the EFOSC2 spectrograph on the ESO 2.2 m telescope. The straight line indicates the position of the slit (PA 85\degr) for the spectrum discussed in the present paper. The slit crosses the spiral arms at about 35\arcsec\ galactocentric distance.}
\label{fig:ngc1358}
\end{figure}

NGC~1358 (Fig.~\ref{fig:ngc1358}) is a barred Sa galaxy hosting an active
galactic nucleus (AGN) at a heliocentric velocity of 4028~\kms\
(Theureau et al. 1998), which is included in the distance-limited sample of
Ulvestad \& Wilson (1989) as a Seyfert~2. It attracted our attention because
of its remarkable circumnuclear kinematics which also allows demonstrating the utility 
of the method introduced here.

Assuming a Hubble constant of 71~\kmsmpc\ (Spergel et al. 2003), the
distance of NGC~1358 amounts to 57~Mpc. Then, 1\arcsec\ corresponds to 275~pc. The
outer optical disk of NGC~1358, as seen in Fig.~\ref{fig:ngc1358}, including the narrow
spiral arms, has a radius of 1\arcmin, or 16.5~kpc. The bar along a
position angle of 130\degr\ has a semi-major axis length of 19\arcsec\ (Gerssen et al.
2003) or 5.2~kpc.

The continuum peak at the nucleus shows a half power half width of 1\farcs3. Dumas et al. (2007) derived, by integral field spectroscopy,
the velocity fields of the ionized gas in the central regions in the emission line of [OIII]~$\lambda$~5007
as well as in the stellar absorption lines (the latter displayed only in a poster).
The stellar velocity field shows a kinematic major axis in a position angle
PA~11.5\degr\ with the North side receding. This is also close to the assumed PA of the line of nodes of the outer disk. The authors derive a symmetric stellar
rotation curve out to $\pm 4 \arcsec$ from the nucleus. The [OIII] velocity field, on
the other hand, shows a kinematic major axis in about PA~90\degr\ with the East side
receding.
Narrow band imaging in [OIII]~$\lambda$~5007 shows a slightly S-shaped emission line
region with a maximum intensity 1\farcs6 from the nucleus, again along a PA of about 130\degr\ (Dumas et al. 2007). This feature shows clearly deviating negative velocities. 
Dumas et al. (2007) suggest the elongated nuclear emission line region to correspond to gaseous streaming due to an inner bar inside the Inner Lindblad Resonance of the outer bar.

An HST/STIS spectrum covering the [SII]~$\lambda$~6716, 6731 lines,
within an aperture size of 0\farcs2 centered on the nucleus, shows double
components with the fainter component blueshifted with
respect to the stronger (Rice et al. 2006). 

\section{Observations}
\begin{figure*}
\resizebox{\hsize}{!}{\includegraphics{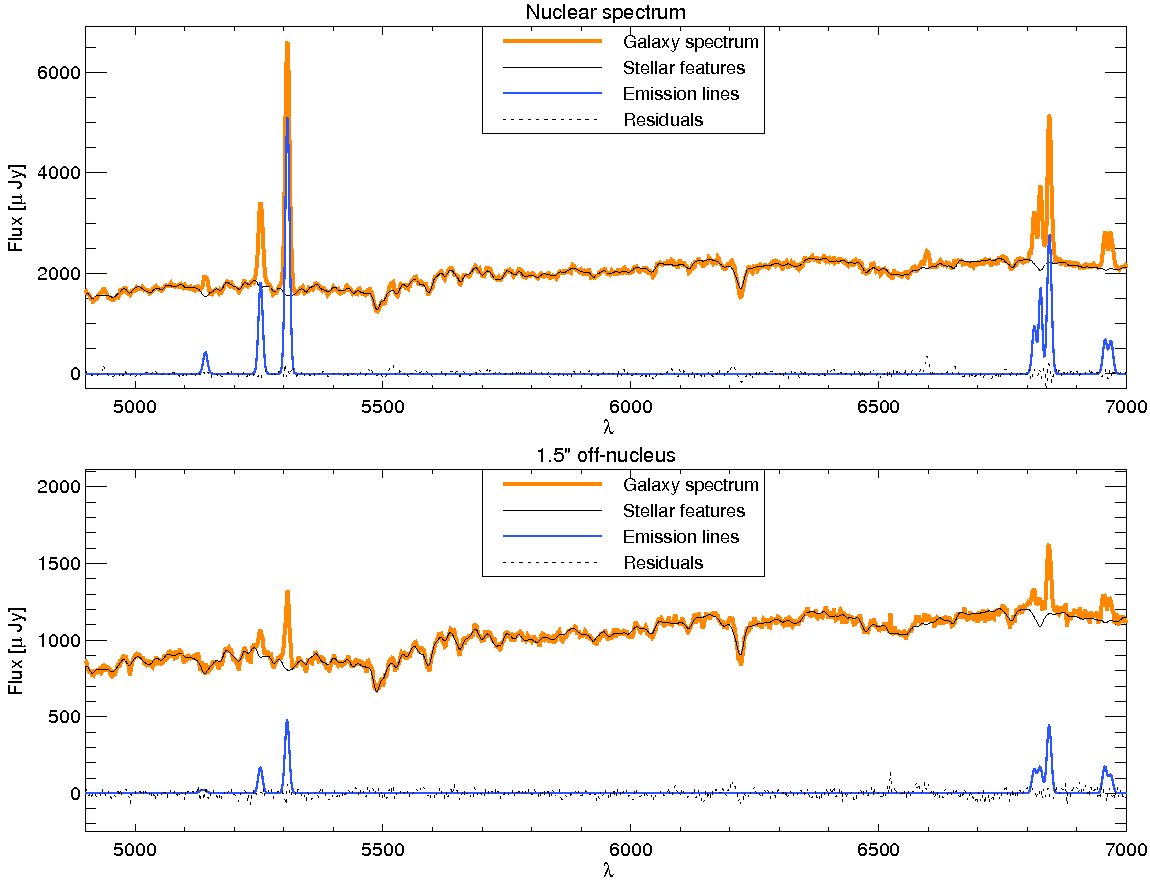}}
\caption{Spectrum of the nucleus (top) and 1\farcs5 West off-nucleus region of NGC~1358. The best fitted linear combination of single stellar populations shows the stellar features fitted simultaneously as  Gaussian fits to the emission lines have been applied. The residuals (data - total fit) show that the fits have been successful ([OI]~$\lambda$~6300, 6364 have been excluded in this fit).}
\label{fig:spectra}
\end{figure*}

We have observed NGC~1358 with the EFOSC2 spectrograph on the ESO 2.2~m telescope and obtained a long slit spectrum over the nucleus in PA~85\degr\ on December 6th 1996. With this position angle a star was brought on the slit, which gave a check on the position angle as well as the PSF. The slit cuts the main spiral arms about 35\arcsec\ East and West of the nucleus (Fig.~\ref{fig:ngc1358}). The spectrum covered the wavelength region 4075 -- 7000 \AA, where the strongest spectral features are the H$\beta~\lambda$~4861, [OIII]~$\lambda$~4960, 5007, H$\alpha~\lambda$~6563, [NII]~$\lambda$~6548, 6584, [SII]~$\lambda$~6716, 6731 emission lines and the Mgb and NaD absorption lines (Fig.~\ref{fig:spectra}). In the spatial direction the pixel size corresponds to 0\farcs26.  As flux calibration star Hiltner 600 was chosen.

Figure~\ref{fig:spectra} shows two fully reduced and flux calibrated example spectra obtained from the long slit spectrum. To derive the distribution and kinematic parameters, we have fitted the galaxy spectrum in two different ways.
Firstly, by using the standard IRAF routine SPLOT to fit Gaussians to the emission and absorption lines, and secondly, by using the simultaneous gas and absorption line fitting IDL routine developed by Sarzi et al. (2006) and Falc\'on-Barroso et al. (2006). The latter method is designed to separate the relative contributions of stellar continuum and absorption features and nebular emission in spectra of galaxies, while measuring the gas emission and kinematics. Accordingly, it uses the routine of Cappellari \& Emsellem (2004) to find the best fitted linear combination of single stellar population spectra by Vazdekis (1999) to build a stellar spectrum representative of the observed galaxy spectrum for this wavelength range and with the same spectral resolution. Simultaneously, the nebular emission lines have been fitted, and the results yield the gas and stellar kinematics. We found that both  routines deliver fully comparable gas distribution and kinematics, with the advantage that the  IDL routine has derived a much more robust stellar kinematics.

\begin{figure*}
\resizebox{\hsize}{!}{\includegraphics{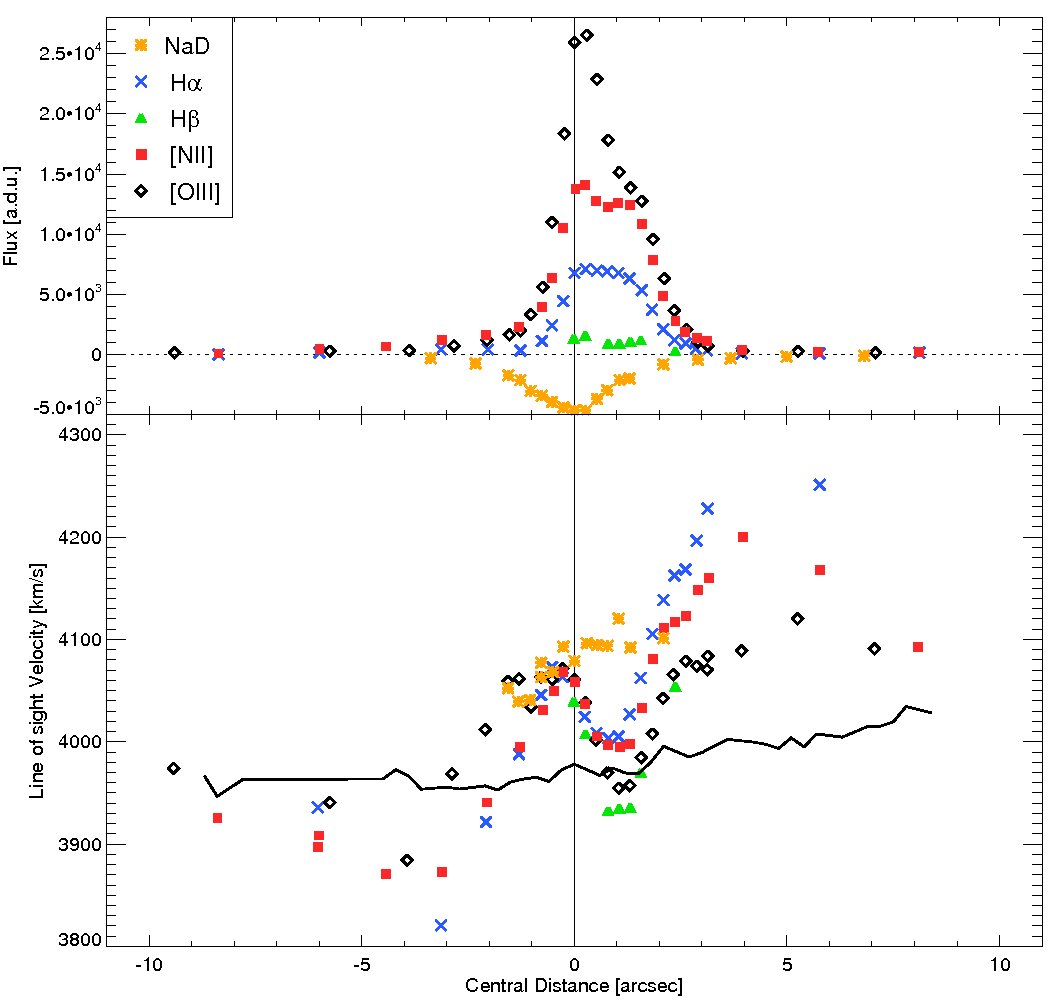}}
\caption{Fluxes and velocities over the central region of NGC~1358 along the slit in PA~85\degr. The flux scale is arbitrary. The solid curve illustrates the velocity of the stellar component as derived from all the stellar features present in the spectra. East is to the right and the zero point of the abscissa corresponds to the maximum of the continuum (Fig.~\ref{fig:continuum}), and the NaD absorption line flux values have been inverted for presentation purposes. We have derived velocity uncertainties of the order of 10 \kms.}
\label{fig:observedemission}
\end{figure*}

Figure~\ref{fig:observedemission} shows the run of [NII]~$\lambda$~6584, H$\alpha$, [OIII]~$\lambda$~5007 and H$\beta$ emission as well as NaD absorption intensities and velocities over the central region of NGC~1358 along the slit as derived with the SPLOT routine. The solid line shows the velocities of the stellar component derived from all the stellar features.
As zero-point of the abscissa we have chosen the position of the maximum of the continuum (cp. Fig.~\ref{fig:continuum}).
In Fig.~\ref{fig:observedemission} the emission line velocities, and in particular H$\alpha$ and [NII], show a rapidly increasing rotation curve with a maximum about 4\arcsec\ from the center. Dumas et al. (2007) show that this nuclear emission line disk should have a kinematic major axis directed close to PA~90\degr, i.e. only 5\degr\ from the PA of the present slit position, but almost perpendicular to the kinematic major axis of the stellar bulge.

Within 2\arcsec\ from the center, on the East side, we see a reversal, or a dip, in the apparent rotation curve of the emission lines. In the NaD absorption line velocities, this dip is absent. From Fig.~\ref{fig:observedemission} it is clear that the emission line velocity dip corresponds to an excess of the emission line intensities up to 2\arcsec\ from the nucleus. The NaD lines show no such asymmetry and the position of the maximum absorption coincides with that of the maximum of continuum emission.

The simplest interpretation of the gas kinematics observed in Fig.~\ref{fig:observedemission} seems to be a central point source and a jet-like outflow that are superimposed on a gaseous disk rotating in the potential of the stellar bulge. Preliminary simulations along the scheme presented below suggested that the jet outflow line of sight velocities should be of the order of $-$200 \kms. 

\begin{figure}
\resizebox{\hsize}{!}{\includegraphics{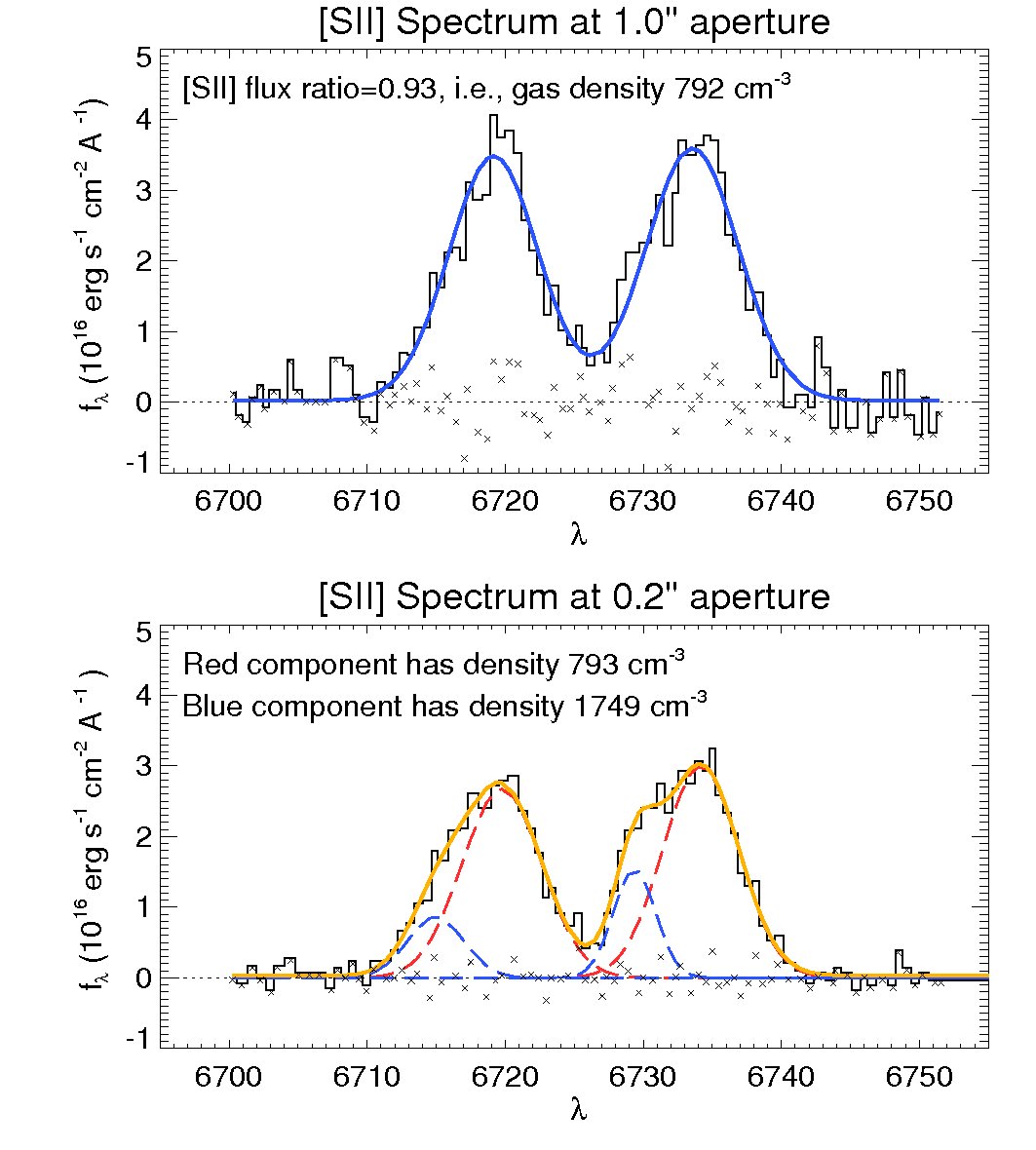}}
\caption{[SII]~$\lambda$ 6716, 6731 spectrum from HST/STIS observations of the nucleus of NGC~1358 (Rice et al. 2006), with constrained Gaussian fits. In the bottom panel, the two components are separated by 214 \kms, and all densities are calculated assuming 10000~K electron temperature. Unconstrained Gaussian fits result in almost identical fits.}
\label{fig:SII}
\end{figure}

A support for this scenario is found in the [SII] emission line profiles from HST/STIS spectra published by Rice et al. (2006) reproduced in Fig.~\ref{fig:SII}. One spectrum is observed through a 1\farcs0 aperture and another through an aperture of 0\farcs2. The applied Gaussian fits have been constrained to enforce the relative separation between the [SII] doublets. In both spectra a strong component is seen in the two lines with very similar flux ratios in the two apertures. In the 0\farcs2 aperture spectrum a secondary blueshifted component is seen. We have assumed 10000~K electron temperature and used the atomic parameters of Mendoza (1983) and Osterbrock (1989) to calculate the electron density from these spectra. The calculated relative fluxes yield for the spatially extended and dominating component an electron density of $793 \,\rm cm^{-3}$. The central component, only appearing in the 0\farcs2 aperture spectrum, has a density of $1749 \,\rm cm^{-3}$. This denser component is fainter and blueshifted with a velocity difference between the strong and faint components of 214 \kms, being the innermost part of the jet, respectively.

The stellar velocities show only a very slight variation over the region in agreement with the slight positive gradient expected for the stellar velocity field of Dumas et al. (2007) with the present slit inclined 74\degr\ to the stellar kinematic major axis. The NaD absorption in front of the nucleus appears to have a higher redshift than the average stellar template redshift, which also is indicated by the shape of the line residuals in Fig.~\ref{fig:spectra}. 

Most remarkably the velocity dip to the East of the nucleus is larger for [OIII] than for the [NII] and H$\alpha$ lines and apparently still larger for H$\beta$.  Outside the dip [OIII] shows smaller rotational velocities than H$\alpha$ and [NII].
Here we ask ourselves, what additional information on the structure and ionization distribution in the nuclear region do these differences in velocity behaviour give us?

\section{Theory}
An explanation for the differences in velocity behaviour between the various emission lines would be that the whole central region consists of different gaseous components with different velocity behaviour, different excitation, and different dust absorption, and thus different sets of line ratios. When this region is smoothed by the PSF, different distortions of the velocity field for the different
lines will result.

Assume  that the region considered covers $N$ pixels and contains $K$ separate structural components, each with its own line ratios and set of velocities. If $J_i^{(\lambda)}$ and $V_i^{(\lambda)}$ are the observed line intensities and velocities at pixel number $i$ of a line with wavelength $\lambda$, we have

\begin{equation}
\label{eq:j}
J_i^{(\lambda)} \, = \, \sum_j \sum_k \, I_{kj}^{(\lambda)} \, p(i-j),
\end{equation}
\begin{equation}
\label{eq:v}
V_i^{(\lambda)} \, = \, \frac{1}{J_i^{(\lambda)}}\displaystyle{\sum_j \sum_k \, v_{kj} \, I_{kj}^{(\lambda)}  \, p(i-j)},
\end{equation}
where $I_{kj}^{(\lambda)}$ is the intrinsic intensity, in the line $\lambda$, of the component $k=(1, \dots, K)$ at the pixel $j$, and $v_{kj}$ the corresponding velocity, which is supposed independent of $\lambda$, and $p(x)$ the PSF. If $L$ is the number of lines measured, we have in the general case $KN(L+1)$ unknown and $2NL$ equations. If the line ratios are all the same in the different components, equation (\ref{eq:v}) is independent of $\lambda$. In the simplest case, where we only consider one component $(K=1)$, we have enough equations to solve for the unknown intensities and velocities for each $\lambda$, which can be made by  CLEAN-like algorithms (H\"ogbom 1974).

In the general case, we do not have a sufficient number of equations to solve for the unknown intensities $I_{kj}^{(\lambda)}$ and velocities $v_{kj}$. If, however, we increase the number of lines measured and at the same time impose some conditions on the unknowns in terms of models, we may in principle be able to reach a solution. For instance, we see from equations (\ref{eq:j}) and (\ref{eq:v}) that if the line ratios $I_{kj}^{(\lambda_1)}/I_{kj}^{(\lambda_2)}$ depend on $k$, i.e., do vary from one component to another, then we have $V_i^{(\lambda_1)}-V_i^{(\lambda_2)} \ne 0$. Thus, the differences between velocities from different spectral lines give us information about differing line ratios between different components. We will illustrate this for our spectrum of NGC~1358.

\section{Models}
We will test velocity and intensity observations over the nuclear region against a scenario where the dip in the rotation curve is caused by an outflow of excited gas from a central source at rest with respect to a nuclear gaseous disk. We strive to have in this model an absolute minimum of free parameters to fit the observations. We will first demonstrate the simplest primitive model, here called Model 1, to motivate why we introduce our more elaborate models.

\subsection{Model 1}
\begin{figure}
\resizebox{\hsize}{!}{\includegraphics{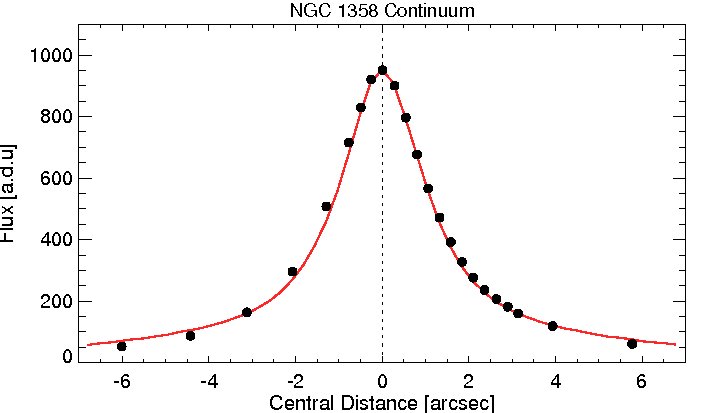}}
\caption{Continuum levels under the emission line H$\alpha$ fitted with the sum of two exponential disks of scale lengths 0\farcs7 and 4\arcsec\ and central intensity ratio 4/1, smoothed with the PSF.}
\label{fig:continuum}
\end{figure}

Our Model 1 has the following features:

\begin{itemize}
\item[a)] A stellar bulge with a mass distribution corresponding to the continuum surface intensity distribution, giving the gravitational field in the nuclear region. As seen in Fig.~\ref{fig:continuum} the observed surface intensity distribution can be well represented by the sum of two exponential disks with scale lengths of 0\farcs7 and 4\arcsec\ respectively and an intensity ratio of 4:1, smoothed with the observed PSF.
To derive a likely potential and rotation curve for this stellar bulge we use the relation (2--170) of Binney \& Tremaine (1987) to give the potential of  the sum of two exponential spheres with scale lengths 0\farcs7 and 4\farcs0 respectively, and a mass ratio of 4:1 in conformance with Fig.~\ref{fig:continuum}. Although we note that routinely bulges are fitted by de Vaucouleurs or Sersic profiles, experience has shown that exponential profiles are  satisfactory to parametrize a bulge (e.g., Fathi \& Peletier 2003, Balcells et al. 2007), and certainly this seems also to be the case for NGC~1358. This makes the total mass $M \times \sin^{2}(i)$ of the bulge the only free parameter of the bulge potential, where $i$ is the 
inclination of the plane of the emission line disk to the plane of the sky 
and we assume that the difference in PA between the slit and the line of 
nodes of the disk is small.

\item[b)]In the potential of this bulge there is a rotating emission line disk. To avoid edge problems we assume this disk to be exponential and rotate with circular velocities. As the intensities of the disk is low compared to the other features, the exact scale length is not crucial, and we fix it ad hoc to 4\farcs0. Then the only free parameters are the central surface line intensities of the emission line disk. 

\item[c)] A central emission line point source coinciding with the center of the bulge and the emission line disk and at rest with respect to that center.

\item[d)] A gas flow, here called the ``jet'', seen from the nucleus out to 7 pixels (1\farcs8) from the nucleus with constant line of sight velocity $- 214$ \kms\ with respect to the point source, which was the velocity of the fainter component relative to the stronger as seen in the HST/STIS spectrum (Fig.~\ref{fig:SII}).
\end{itemize}

\begin{figure}
\resizebox{\hsize}{!}{\includegraphics{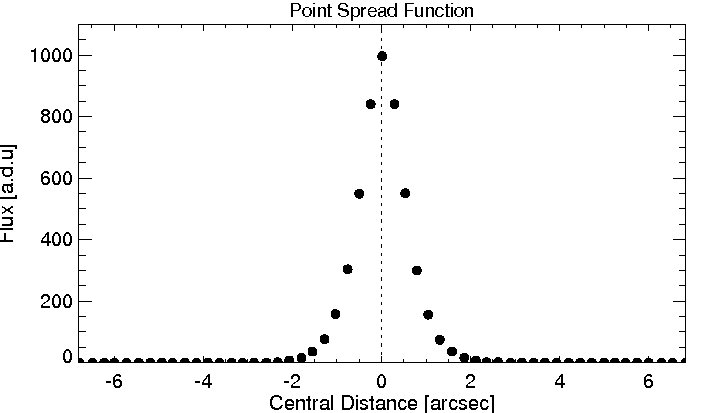}}
\caption{The PSF as derived from the star 150\arcsec\ West of NGC~1358.}
\label{fig:psf}
\end{figure}

The resulting intensities and velocities are convolved with the observed PSF as derived from a field star on the slit (Fig.~\ref{fig:psf}).

The line of sight velocity, or redshift, of the central source and disk is an open question. Theureau et al. (1998) report a redshift of 4028 \kms, while  Dumas et al. (2007) estimate the systemic redshift of the galaxy at $3997 \pm 13$ \kms. Rice et al. (2006) assume a systematic redshift of 4008 \kms, and their spectrum (Fig.~\ref{fig:SII}) shows a line of sight velocity of $+ 145$ \kms\ for the stronger nuclear component with respect to the 4008 \kms\ systemic velocity, i.e. a nuclear redshift of 4153 \kms\ for the stronger component and 3939 \kms\ for the blueshifted weaker component. We prefer to make the systemic velocity of the nuclear disk and the central emission source a free parameter in our model. The additional free parameters are the intensities of the central source and along the jet for each emission line.

The fit of this preliminary model is given in Fig.~\ref{fig:badmodel}. As can be seen, this model cannot reproduce the very sharp rise of the emission line velocity curves at the end of the jet about 2\arcsec\ to the East of the center. Thus, we feel forced to introduce a sharp shock at the end of the jet and a postshock cloud.

\subsection{Model 2}
This model has the following features:

\begin{itemize}
\item[a)] A stellar bulge identical with the bulge in Model 1 presented above, where the total mass $M \times \sin^{2}(i)$ of the bulge is the only free parameter.

\item[b)]In the potential of this bulge a rotating  exponential emission line disk similar to the one in Model 1, again with a scale length of 4\farcs0  and the only free parameters are the central surface line intensities. 

\item[c)] A central emission line point source coinciding with the centra of the bulge and the emission line disk and at rest with respect to that center.

\item[d)] A gas flow, here called the ``jet'', seen from the nucleus out to a certain number of pixels from the nucleus, with constant line of sight velocity $- 214$ \kms\ with respect to the point source. This again is the velocity of the fainter component relative to the stronger as seen in the HST/STIS spectrum (Fig.~\ref{fig:SII}). In a chosen pixel at the end of the jet, we introduce a sharp shock, with a jump to more positive velocities, and in the following pixels, up to pixel 8, a postshock cloud with constant velocity. The mean velocity in the shock pixel is given some intermediate velocity between those of the jet and the postshock cloud.
\end{itemize}

The resulting intensities and velocities are convolved with the observed PSF as derived from a field star on the slit (Fig.~\ref{fig:psf}).

Still, in this way we have tried to keep the free parameters to a minimum, and the free parameters of this model, to be fitted to the observations, are:

\begin{enumerate}
\item The total mass $M \times \sin^{2}(i)$ of the bulge.
\item The common redshift of the central point source and the center of the emission line disk.
\item The intensity of the central point source for each emission line.
\item The central intensity of the emission line disk for each emission line.
\item The pixel to contain the shock and the intensity and velocity in that pixel.
\item The intensities along the jet flow for each emission line.
\item The common velocity and the intensities in the postshock cloud for each emission line.
\end{enumerate}

These parameters are varied until a reasonable fit to the observed velocities and intensities have been obtained. The fit is displayed in Fig.~\ref{fig:Model2}. The shock is placed within pixel 5 with the intermediate velocity of $-100$ \kms. This determines the postshock cloud velocity to about +100 \kms. As can be seen in Fig.~\ref{fig:Model2}, the line intensities of [OIII] and the velocities in the jet are not perfectly fitted, and to place the shock short of pixel 5 makes the situation worse.

Further, it can be seen from the observations that the emission lines have long faint wings on the Western (negative) side, actually H$\alpha$ and [NII] stretching furthest and [OIII] having the shortest wing. The run of the H$\alpha$ and [NII] velocities on the West side show that the main intensities of this extended emission come from the rotating disk. This indicates an asymmetry in the emission line disk, and this makes it difficult with the present model to fit the velocities away from the center simultaneously on both sides, as can be seen in Fig.~\ref{fig:Model2}. Due to the jet, the observed velocities on the left side are more influenced by the inner part of the disk, and on the right side by the outer part. We can then introduce an asymmetry of the observations without abandoning the circular symmetry of the disk by describing its surface intensity as the sum of two superposed exponential disks with widely different scale lengths and varying their central intensities independently. This results in differences in the line ratios, simulating a gradient of these ratios within the disk. This leads us to the next model.

\subsection{Model 3}
This model is the final one which gives us the best fit to the observed velocities and intensities in the nuclear region of the galaxy. Its features are:

\begin{itemize}
\item[a)] A stellar bulge identical with the bulge in Model 1 presented above, where the total mass $M \times \sin^{2}(i)$ of the bulge is the only free parameter.

\item[b)]In the potential of this bulge there is a rotating emission line disk. We assume the intensities in the disk to be the sum of two exponential disks with differing scale lengths. As the intensities of the disk are low compared to the other features, the exact scale lengths are not crucial, and we fix them ad hoc to 0\farcs7 and 4\farcs0 (only for historical reasons the same as for the bulge). Then the only free parameters are the central surface line intensities of the two disks (here called Disk~1 and Disk~2) representing the inner and outer part respectively of the emission line disk.

\item[c)] A central emission line point source coinciding with the centra of the bulge and the emission line disk and at rest with respect to that center.

\item[d)] A gas flow, here called the ``jet'', seen from the nucleus out to a certain number of pixels from the nucleus, with constant line of sight velocity $- 214$ \kms\ with respect to the point source. This again is the velocity of the fainter component relative to the stronger as seen in the HST/STIS spectrum (Fig.~\ref{fig:SII}). At the end of this jet, at the border between two pixels, we introduce a sharp shock, with a jump to more positive velocities, and in the following pixels, up to pixel 7, a postshock cloud with constant velocity. 
\end{itemize}
The resulting intensities and velocities are convolved with the observed PSF as derived from the field star on the slit (Fig.~\ref{fig:psf}).

The number of free parameters of this model is less than that of Model 2 and they are:

\begin{enumerate}
\item The total mass $M \times \sin^{2}(i)$ of the bulge.
\item The common redshift of the central point source and the center of the emission line disk.
\item The intensity of the central point source for each emission line.
\item The central intensities of the two superimposed emission disks for each emission line. 
\item The position of the last pixel of the jet.
\item The intensities along the jet flow for each emission line.
\item The common velocity and the intensities in the postshock cloud for each emission line.
\end{enumerate}

\begin{figure*}
\resizebox{\hsize}{!}{\includegraphics{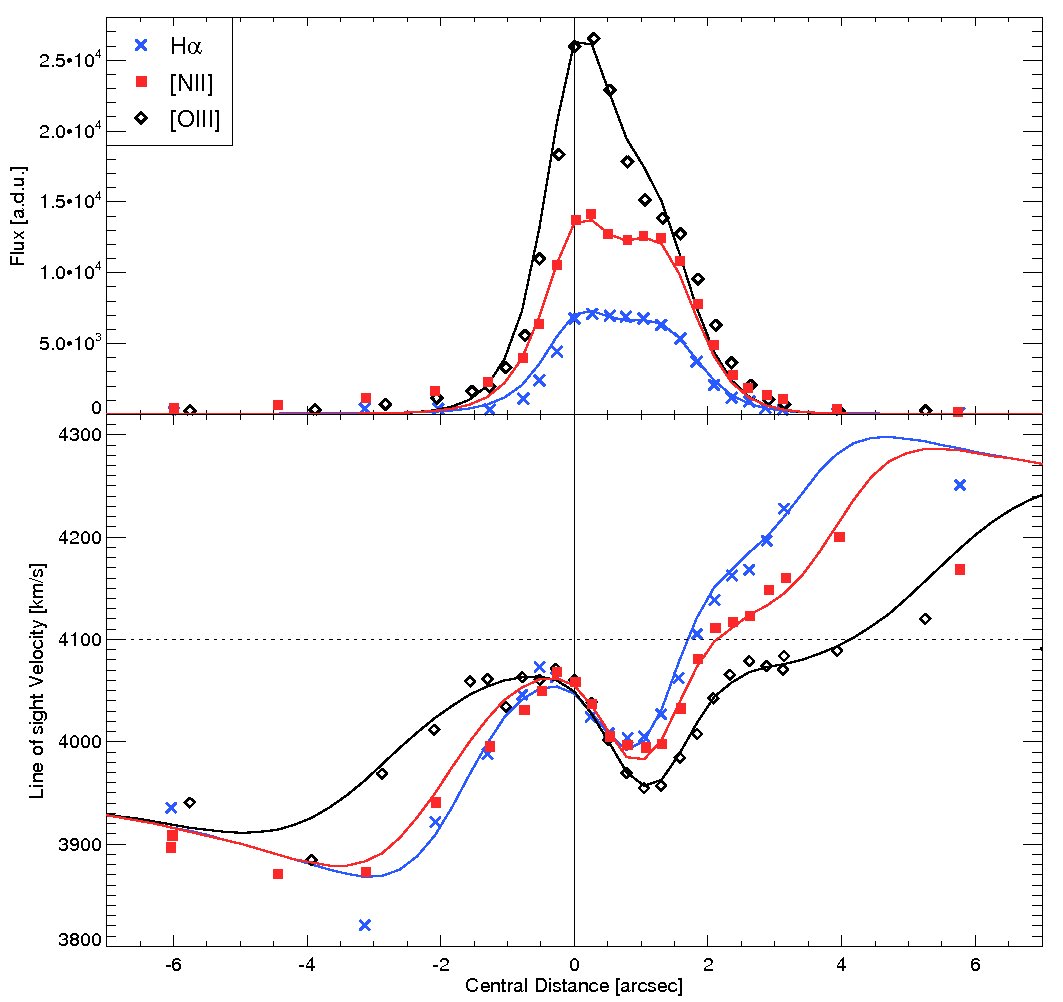}}
\caption{Best model fit to the emission line intensities and line of sight velocities (Model~3 as described in the text). The dotted line indicates the derived redshift of the nuclear source.}
\label{fig:model}
\end{figure*}

The fit of Model 3 to the velocities and intensities of the various emission lines is displayed in Fig.~\ref{fig:model}, and the parameters are given in Table~\ref{tab:modelvalues}. The table also contains measurements for the spiral arms with the SPLOT routine where they intersect with the present slit.

\begin{table*}
\caption{Parameters for the model presented in Fig.~\ref{fig:model}. Disk~1 and Disk~2 make up the emission line disk with scale lengths $r_d$ of 0\farcs7 and 4\arcsec\ respectively (c.f. Fig.~\ref{fig:tableplot}).}
\centering \tabcolsep=3pt
\begin{tabular}{l|r|rrrrr|rr|rr|rr}
\hline \hline
\multicolumn{1}{c}{Component}&
\multicolumn{1}{c}{Nucleus}&
\multicolumn{5}{c}{$\longleftarrow\;$ Jet $\;\longrightarrow$}&
\multicolumn{2}{c}{$\longleftarrow\;$ Cloud $\;\longrightarrow$}&
\multicolumn{1}{c}{Disk 1}&
\multicolumn{1}{c}{Disk 2}&
\multicolumn{2}{c}{$\longleftarrow\;$ Spiral arms $\;\longrightarrow$}\\
\hline
\hline
Pixel        &  0  &    1 &   2  &   3   &  4   &   5  &   6  &   7  &  \    &   \     &   128 & $-140$\\
Arcsec       &  0  & 0.26 & 0.52 & 0.78  & 1.05 &  1.31&  1.6 &  1.8 &   $r_d = 0.7$  &    $r_d = 4$    &   33  & $-37$\\
Velocity \kms&0    &$-214$&$-214$&$-214$ &$-214$&$-214$&$+250$&$+250$& Rotating & Rotating& $-58$     & $-116$\\
Flux (H$\alpha$)&4700 &  910 &  910 &  450  & 1650 &  2100&  1700&  310 &    116   &    29   &   210     &  40\\
Flux ([NII]) &9600 & 1400 & 1140 & 1100  & 2700 &  5100&  2600&  310 &    250   &    12 &    72     &  24\\
Flux ([OIII]) &19400& 3900 & 2600 & 1720  & 3900 &  6000&  1500&  600 &    160   &     0 &    \      &    \\
\multicolumn{1}{l|}{log ([NII]/H$\alpha$)}&
\multicolumn{1}{r|}{0.31}&
\multicolumn{5}{c|}{$\longleftarrow\;$ 0.28 $\;\longrightarrow$}&
\multicolumn{2}{c|}{$\longleftarrow\;$ 0.16 $\;\longrightarrow$}&
\multicolumn{1}{r}{0.33}&
\multicolumn{1}{r|}{$-0.38$}&
\multicolumn{1}{r}{$-0.46$}&
\multicolumn{1}{r}{$-0.22$}\\
\multicolumn{1}{l|}{log ([OIII]/H$\alpha$)}&
\multicolumn{1}{r|}{0.62}&
\multicolumn{5}{c|}{$\longleftarrow\;$ 0.48 $\;\longrightarrow$}&
\multicolumn{2}{c|}{$\longleftarrow\;$ 0.02 $\;\longrightarrow$}&
\multicolumn{1}{r}{0.14}&
\multicolumn{1}{r|}{\ }&
\multicolumn{1}{r}{\ }\\
\hline
\end{tabular}
\label{tab:modelvalues}
\end{table*}

\section{Results}
\begin{figure}
\resizebox{\hsize}{!}{\includegraphics{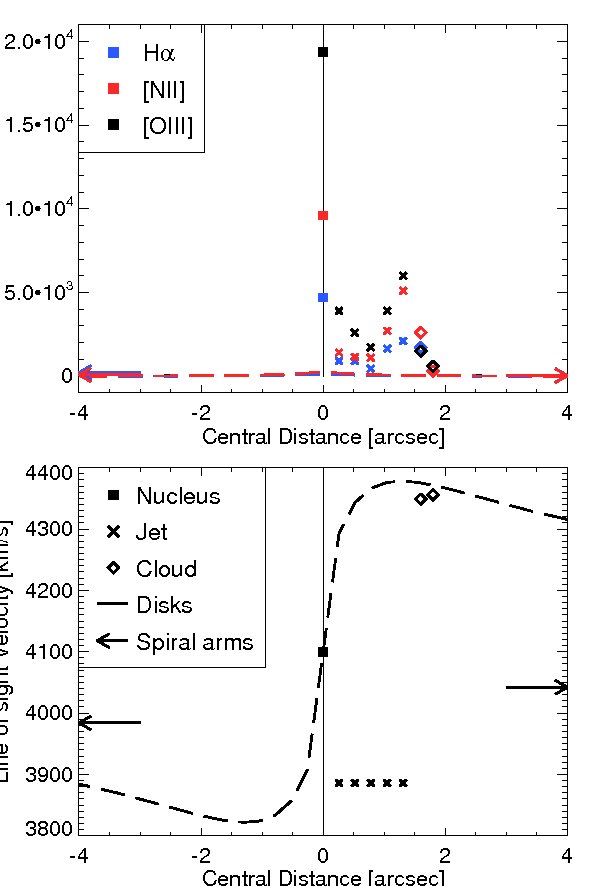}}
\caption{Illustrating fluxes and velocities of the individual components listed in Table~\ref{tab:modelvalues} without the effect of the PSF. The arrows show the fluxes and velocities assigned to the spiral arms which are fixed at -37\arcsec\ and 33\arcsec\ galactocentric distance. Filled boxes illustrate the nucleus, crosses the jet, and diamonds the cloud, respectively.}
\label{fig:tableplot}
\end{figure}

Our best fit, Model 3, gives an observed redshift of the central source of 4100 \kms, or 4089 \kms\ with heliocentric velocity correction. The total $M \times \sin^{2}(i)$ of the bulge is determined essentially by the velocities of H$\alpha$ and [NII] at the distance $-6$\arcsec\ from the center. The best fit gives $M \times \sin^{2}(i) = 14 \times 10^{9}~\rm M_{\sun}$. The remaining parameters of this fit are given in Table~\ref{tab:modelvalues}. Fig.~\ref{fig:tableplot} illustrates Table~\ref{tab:modelvalues} and gives the intensities and velocities of the various features before smearing with the PSF.

The central intensities of the emission line disks are essentially given by the run of velocities from $-3\arcsec$ to $-1\arcsec$ and from $+2\arcsec$ to $+4\arcsec$ from the center. 

The position of the shock falls at the border between pixel 5 and 6, i.e. 1\farcs5, from the center. This, in turn, forces the postshock velocity up to about +250 \kms in pixels 6 and 7. 

Figure~\ref{fig:model} shows that the velocities and intensities can be reproduced with fair accuracy within this nuclear region. Finally, the diagnostic line ratios log~[NII]/H$\alpha$ and log~[OIII]/H$\alpha$, are given in Table~\ref{tab:modelvalues} for the various components. The H$\beta$ intensities are too weak and uncertain to give a unique solution in the present case. 

\section{Discussion}
Our ``best fit'' Model~3 resolves the nuclear structure of NGC~1358 into

\begin{itemize}
\item[1)]a central unresolved emission line source (the Nucleus),
\item[2)]a 1\farcs5 long jet emerging from the nucleus with a line of sight velocity of $-214$ \kms,
\item[3)]a spherical nuclear stellar bulge containing a rotating emission line disk, inclined to the stellar kinematic symmetry plane of the bulge,
\item[4)]an emission line region, the postshock cloud, at the end of the jet, and after a drastic jump in velocity.
\end{itemize}

\begin{figure}
\resizebox{\hsize}{!}{\includegraphics{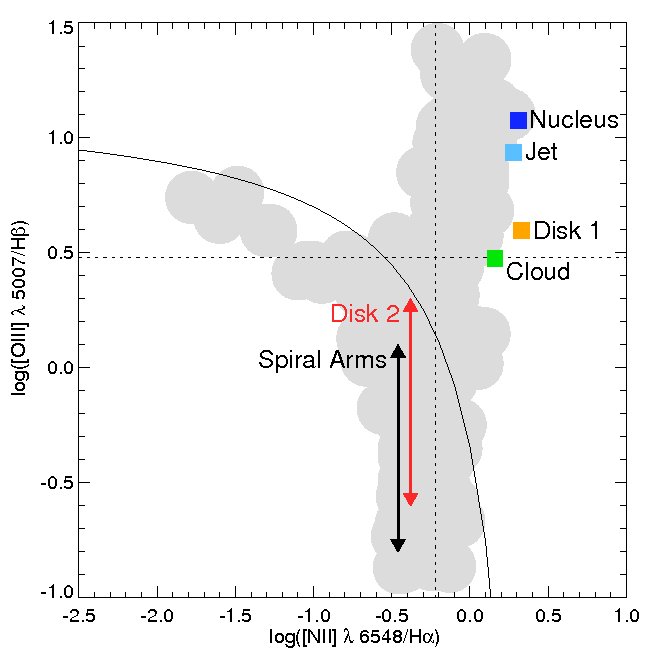}}
\caption{Line ratio diagnostics derived from our ESO spectra. The solid curve shows the Starburst-AGN separation lines of Kewley et al. (2001), with the shaded region indicating their sample galaxies. }
\label{fig:lineratios}
\end{figure}

With the emission line fluxes for these structures we are able to set up the Starburst-AGN diagnostic diagrams introduced by Baldwin, Phillips \& Terlevich (1981) and Veilleux \& Osterbrock (1987). Figure~\ref{fig:lineratios} shows the diagnostic diagram log~[OIII]/H$\beta$ versus log~[NII]/H$\alpha$ where the [OIII]/H$\beta$ ratio is derived from [OIII]/H$\alpha$ assuming large optical depth, referred to as Case B described in chapter 4.2 of Osterbrock (1989). Accordingly, for an electron temperature of 10000 K, the H$\alpha$/H$\beta$ is 2.87, which we note is not much different from the case when the gas is assumed to be optically thin. As evident, the nuclear source and the jet fall in the region of highly excited AGNs, while the inner disk and the postshock cloud fall close to the region of Low Ionization Nuclear Emission line Regions (LINERs). The ratios [NII]/H$\alpha$ places the outer part of the nuclear disk and the spiral arms in the HII region domain.

The contribution from the underlying H$\alpha$ and H$\beta$ absorption lines only change the derived emission line ratios marginally. However, the effect of reddening has not been taken into account in the emission ratios presented in Fig.~\ref{fig:lineratios} and Table~\ref{tab:modelvalues}. As we have used the unreddened ratio H$\alpha$/H$\beta$, our derived [OIII]/H$\beta$ ratios presented in the diagram are lower limits in the presence of reddening. This is an effect that favours our conclusions about the nature of the components.

The nuclear point source shows the highest excitation. If H$\beta$ would be included in this analysis, its derived true intensity in the Nuclear point source would with H$\alpha$ give the true absorption of the nuclear source of this Seyfert~2 galaxy. Unfortunately, H$\beta$ is much too faint and uncertain to be fitted with our procedure. Stronger exposed H$\beta$ would be needed.

The emission line map of Dumas et al. (2007) shows that the maximum intensity for our jet lies in about PA 130\degr, and there is also a fainter counter-jet in the opposite direction. Thus, our slit avoids the maximum intensity of the jet and also the counter-jet. As seen from Fig.~\ref{fig:model}, this means that there is no need for a counter-jet in our model. However, Disk~1 which is completely drowned on he East side by the jet, may in reality correspond to a trace of an absorbed counter-jet. Its position fin Fig.~\ref{fig:lineratios} does not contradict this. On the other hand, the introduction of a counter-jet in our model would add a number of uncertain intensity parameters.

The HST/STIS spectrum (Fig.~\ref{fig:SII}) indicates a high density for the jet, and the jet intensity maps display a bent spiral-like morphology consistent with a precessing jet flow. Considering the angle of about 45\degr\ between the jet axis and our slit, the jet may be surrounded by a more widespread outflow, similar to the one observed in the nucleus of the Seyfert~1.5 galaxy NGC~1365 (Hjelm \& Lindblad 1996). Moran et al. (2000) conclude 
that NGC~1358 belongs to the class of about 50~\% Seyfert~2s that do not 
contain a polarized broad line region (NPBL~Sy2). Shu et al. (2007) argue that NPBL Sy2s can be interpreted in the framework of the AGN unified model if the absorption column density is higher at large inclinations and the scattering region is absorbed at large inclinations. Thus, the inclination of the jet to the line of sight may be large and the outflow motion considerably higher than 214 \kms.

The emission line disk intensities are completely hidden behind the strong features in the central part and too faint to be measured at larger distances from the center. They are therefore completely determined by the influence of this disk on the observed total velocities. Disks 1 and 2 represent the inner and outer parts, respectively, of this nuclear emission line disk.
Their line ratios then indicate a rather strong gradient within the nuclear disk (see Table~\ref{tab:modelvalues}), or, as mentioned, a difference in line ratios between an eventual absorbed counter-jet and Disk~2.

The disk rotates in the potential of the stellar bulge at an angle to the kinematic symmetry plane of the bulge. Our slit is oriented close to the kinematic major axis of the disk, but close to the kinematic minor axis of the stellar bulge (Dumas et al. 2007). The redshift of the central source in our model and the systematic redshift of the stellar bulge differ with about 100 \kms (Fig.~\ref{fig:observedemission}). The slit crosses the spiral arms close to the photometric minor axis of the outer galactic disk (Fig.~\ref{fig:ngc1358}. Their mean velocity relative the nucleus is $-87$ \kms.
In view of the fact that our slit is parallel to the line of nodes of the emission line velocity field and perpendicular to the line of nodes of the stellar velocity field, the difference in central velocity between our models and that of the stellar bulge, as seen in Fig.~\ref{fig:observedemission}, may be accounted for if our slit misses the central peak passing about 1\arcsec\ to the South of the nucleus. This would mean that our derived rotation curve is somewhat flatter than the true one and the intensities of the central source too faint. The NaD absorption lines, measured with SPLOT, show the same systematic redshift as the emission lines, which would imply that their main absorption comes from the emission line disk and not from the stellar bulge. The negative mean velocity of the spiral arms speaks somewhat against this miscentering view with whatever weight this may have.

\section{Conclusions}
We want to stress that our method, i.a. because of the necessity to introduce number of more or less reasonable assumptions in order to decrease the number of free parameters, does not give a unique solution to the nuclear structure of NGC 1358. The differing velocities in the different lines give strong constraints for the line ratios. Experiments with our model, for example, shows that it is the weakness of [OIII] in the postshock cloud and nuclear disk that causes the dip in velocity of the Eastern part of the velocity curve to be deeper in [OIII], and the [OIII] velocity curve to be flatter on both sides. But, it is necessary to know with good precision the outer wings of the PSF in order to account for the influence of the strong components on the fainter ones. 

In our particular case the main draw back is that the observations are restricted to a single slit. To get all necessary information about the velocities and spatial distribution of the different components from an isolated spectrum is futile, because of the very limited coverage and the unknown influence due to the point spread function from sources outside the slit. By adding more slits, there is still a difficulty to obtain an accurate and complete coverage (Lindblad et al. 1996). Obviously, the ideal is an integral field spectrometer and an extension of the method to a two-dimensional treatment, combined with comprehensive statistical analysis, covering the entire region of interest.
 
Nevertheless, we argue that, with this analysis, we have made a step towards a resolution of the nuclear region of NGC~1358 into a number of different components with different velocity behaviour and excitation, where a summary of the results are seen in Fig.~\ref{fig:lineratios} and Table~\ref{tab:modelvalues}. 

\section*{Acknowledgments}
We are indebted to the night assistants L.~Ramirez and M.~Turatto for their efficient assistance at the ESO telescope. IRAF is distributed by the National Optical Astronomy Observatories, which are operated by the Association of Universities for Research in Astronomy, Inc., under cooperative agreement with the National Science Foundation. KF acknowledges support from the Swedish Research Council (Vetenskapsr\aa det). Finally, we are grateful to an anonymous referee for valuable suggestions and remarks.

\bibliographystyle{plain}

\appendix
\section{Alternative models}
\begin{figure}
\resizebox{\hsize}{!}{\includegraphics{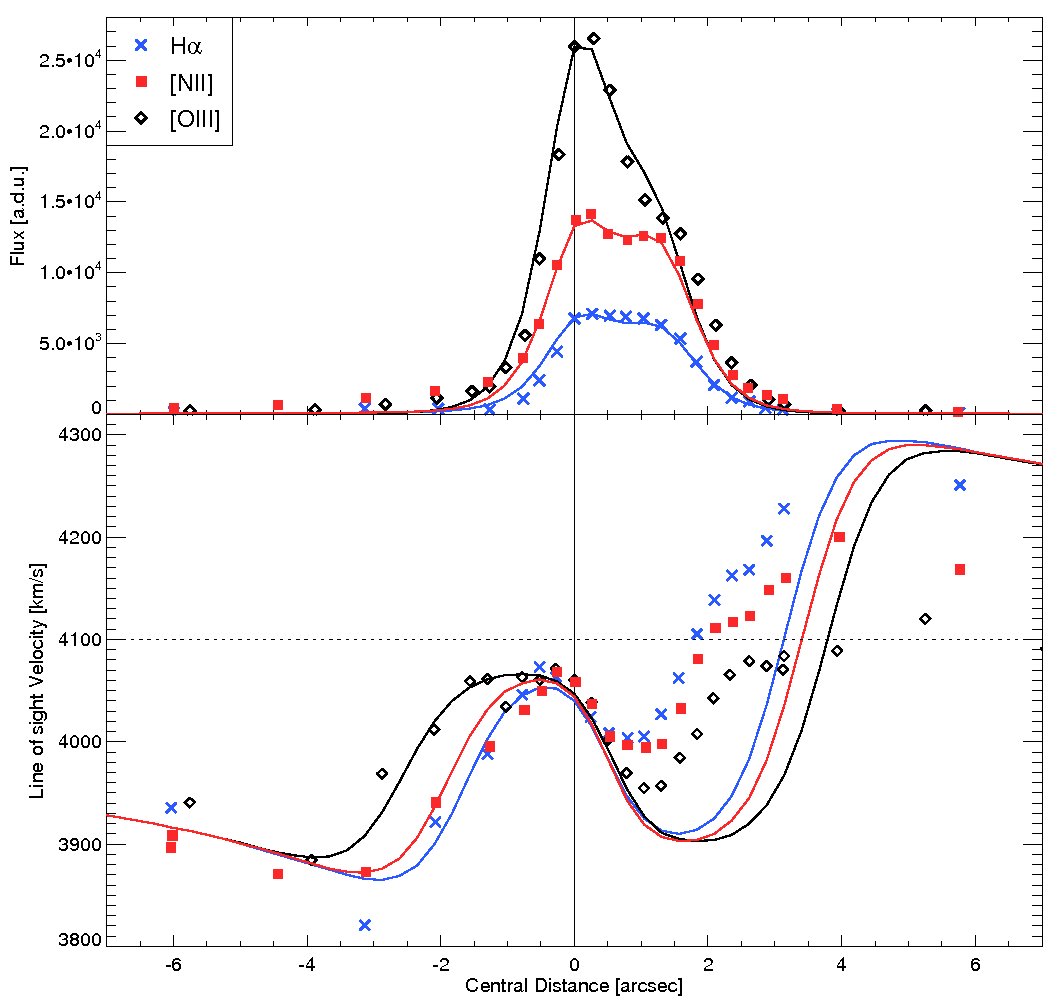}}
\caption{Fit of Model~1 to the observations.}
\label{fig:badmodel}
\end{figure}

\begin{figure}
\resizebox{\hsize}{!}{\includegraphics{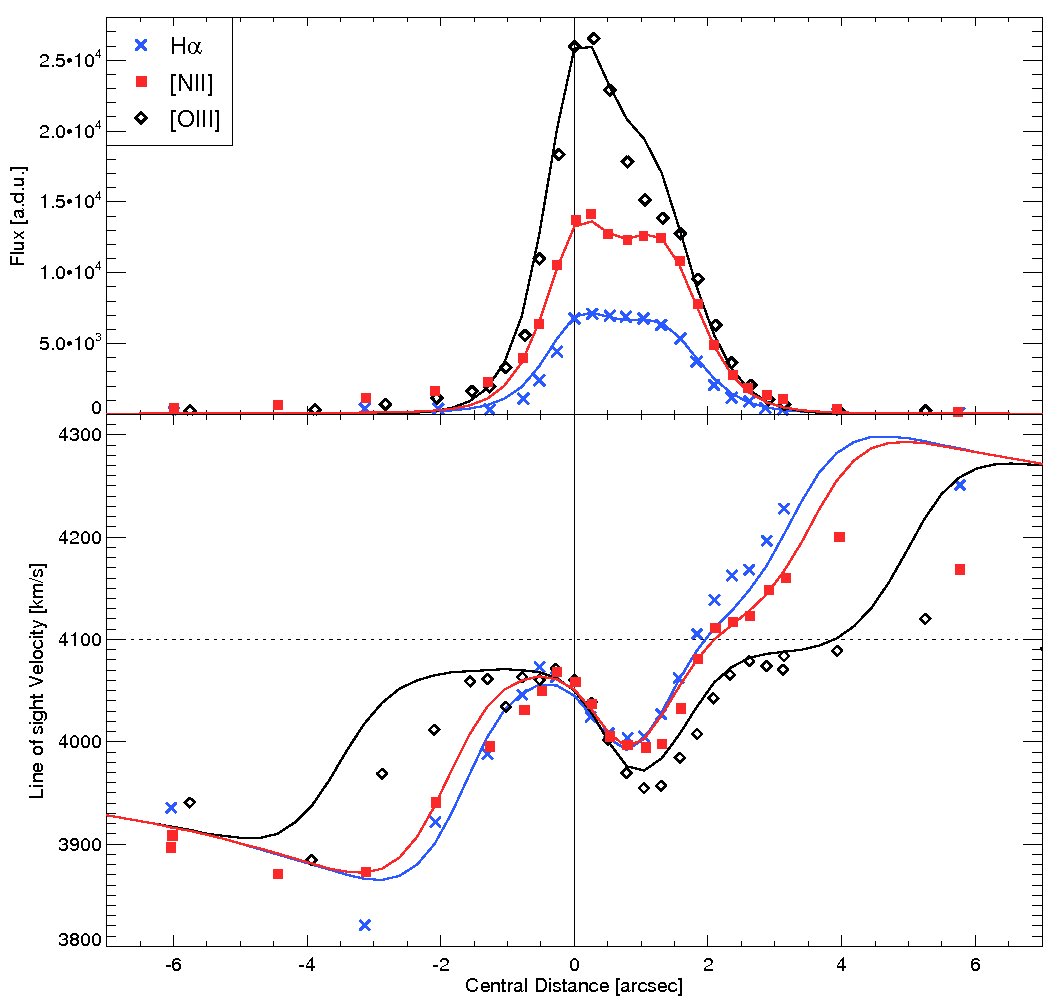}}
\caption{Fit of Model~2 to the observations.}
\label{fig:Model2}
\end{figure}

The model presented in this paper have been obtained after a number of iterations fixing the different free parameters. Here we show two models based on one single exponential emission line disk, both reproducing the observations to different degrees of satisfaction (Fig.~\ref{fig:badmodel} showing Model~1, and Fig.~\ref{fig:Model2} showing Model~2 as described in the text). Although the second model could be seen somewhat comparable to that shown in Fig.~\ref{fig:model}, we chose Fig.~\ref{fig:model} as our preferred model since it reproduces most of the observed flux and kinematic features of NGC~1358.

\end{document}